\documentclass[12pt]{article}
\usepackage{graphics}
\usepackage{amsmath}
\usepackage{amssymb}
\usepackage{graphicx}
\usepackage{makeidx}
\usepackage{mathrsfs}
\usepackage{amsfonts}
\usepackage[mathscr]{eucal}
\usepackage{amsmath}

\usepackage{graphics}

\DeclareGraphicsExtensions{.eps,.pdf,.jpg,.png}

\def\e{{\rm e}}
\def\e{\hbox{e}}
\def\d{\hbox {d}}
\def\dt{\hbox{dt}}
\def\ds{\displaystyle}
\def\RR{\vbox {\hbox to 8.9pt {I\hskip-2.1pt R\hfil}}}
\def\NN{{\rm I\hskip-2pt N}}
\def\CC{{\rm C\hskip-4.8pt \vrule height 6pt width 12000sp\hskip 5pt}}
\def\ni{\noindent}
\def\q{\quad}  \def\qq{\qquad}
\def\cen{\centerline}
\def \rec#1{{\frac{1}{#1}}}
\def\pni{\par\noindent}
\def\vsh{\smallskip}
\def\vs{\medskip}
\def\vvs{\bigskip}
\def\vsp{\vsh\pni} 

\DeclareGraphicsExtensions{.eps,.pdf,.jpg,.png}

\begin{document}

\cen{{\bf FRACALMO PRE-PRINT: \ http://www.fracalmo.org}}

\hrule
\vskip 0.50truecm
\font\title=cmbx12 scaled\magstep2
\font\bfs=cmbx12 scaled\magstep1
\font\little=cmr10
\begin{center}
{\title Becker and Lomnitz rheological models:  }
 \\[0.25truecm]
 {\title a comparison\footnote{%
Paper published in  A. D'Amore, L. Grassia and  D. Acierno (Editors), 
AIP (American Institute of Physics)  Conf. Proc. Vol. 1459, pp. 132-135.
 (ISBN 978-0-7354-1061-9):
  Proceedings of  the International Conference TOP (Times of Polymers \& Composites), Ischia, Italy, 10-14 June 2012.}}
 \\  [0.25truecm]
 Francesco MAINARDI$^{(1)}$ and
Giorgio SPADA$^{(2)}$
\\  [0.25truecm]
$\null^{(1)}$ {\little Department of Physics, University of Bologna, and INFN}
\\
{\little Via Irnerio 46, I-40126 Bologna, Italy}
\\{\little Corresponding Author.   E-mail: francesco.mainardi@bo.infn.it}
\\[0.25truecm]
$\null^{(2)}$ {\little Dipartimento di Scienze di Base e Fondamenti,
 University of Urbino,}
 \\ {\little Via Santa Chiara 27, I-61029 Urbino, Italy}
 \\ {\little E-mail: giorgio.spada@gmail.com}
\end{center}

%
\vsp{\it 2010 Physics and Astronomy Classification Scheme (PACS)}:
46.35.+z, 62.20.Hg, 76.60.-k, 83.60.Df, 91.32.-m
\vsp


\def\e{\hbox{e}}
\def\d{\hbox {d}}
\def\dt{\hbox{dt}}
\def\ds{\displaystyle}
\def\RR{\vbox {\hbox to 8.9pt {I\hskip-2.1pt R\hfil}}}
\def\NN{{\rm I\hskip-2pt N}}
\def\CC{{\rm C\hskip-4.8pt \vrule height 6pt width 12000sp\hskip 5pt}}
\def\ni{\noindent}
\def\q{\quad}  \def\qq{\qquad}
\def\cen{\centerline}
\def \rec#1{{\frac{1}{#1}}}
\def\pni{\par\noindent}
\def\vsh{\smallskip}
\def\vs{\medskip}
\def\vvs{\bigskip}

\begin{abstract}
The viscoelastic material functions for the Becker and the Lomnitz rheological models, 
sometimes employed to describe the transient flow of rocks, are studied and compared. 
Their creep functions, which are known in a closed form, share a similar time dependence and asymptotic behavior. 
This is also found for the relaxation functions, obtained by solving numerically 
a Volterra equation of the second kind. 
We show that the two rheologies constitute a clear example of broadly similar creep and relaxation patterns 
associated with neatly distinct retardation spectra, for which analytical expressions are available. 
\end{abstract}


\section{Introduction}
The purpose of this paper is to draw the attention of polymer scientists on 
two models used in Earth rheology. They  are usually refereed to as Becker and Lomnitz 
to honor the scientists who have introduced them in 1925 \cite{Becker_ZP25} and in 1956
\cite{Lomnitz_JG56}, respectively.
Both models  exhibit  slow varying creep laws suitable for simulating the flow and the 
(quasi frequency independent) energy dissipation  in rocks, see e.g. Strick and Mainardi
\cite{Strick-Mainardi_GJRAS82}. 
Though the corresponding relaxation laws are not considered in geophysical frameworks,  
they are certainly of interest in the  theory and applications of linear viscoelasticty.
As far as we know, in both classical
 and contemporary  polymer science, see e.g. \cite{Ferry,Tschoegel,Grassia-DAmore_2009a,Grassia-DAmore_2009b},  
  these rheological models have not been taken into  account.
   
In the following we will discuss the analytical creep laws for the
two models along with their graphical representation versus dimensionless time both 
in linear and logarithmic scales. 
Because the differences between the two creep laws remain small
as time is evolving, we also show the rate of creep in order to have a better insight of the comparison.
Then, we  numerically compute and visualize the corresponding relaxation laws by solving a Volterra 
integral equation of the second kind.
The major difference between the two models is found  in their retardation spectra.

\section{The creep laws}
In Earth rheology, the law of creep is usually written as
 $$
 J(t)= J_U [1 + q \psi(t)] \,, \quad  t\ge 0\,,
 \eqno(1)$$
 where $t$ is time, $J_U$  is the un-relaxed compliance, $q$ is a  positive dimensionless material constant, and
 $\psi(t)$ is the dimensionless creep function.
 Consistently with the general theory of linear viscoelasticity, $\psi(t)$
is a Bernstein function, that is positive with a completely monotone
  derivative, with a related spectrum of retardation times (see e.g.  Mainardi
  \cite{Mainardi_BOOK10}). 
  
  For the \underline{Becker model} \cite{Becker_ZP25} we have
  $$ \psi^B(t)= \hbox{Ein} (t/\tau _0) \,,
  \quad t \ge 0 \,, \quad  \tau _0>0\,,  \eqno(2)$$
where Ein  denotes the 
{\it modified exponential integral function} (see e.g. \cite{Mainardi_BOOK10}).
Assuming $\tau_0=1$,    
we have the  integral and series representations 
$$ \hbox{Ein}(t) =  \int_{0}^{t}
 \frac{\;1-\e^{\ds -u }}{u }\,du  =
\sum_{n=1}^{\infty}
   (-1)^{n-1}  \frac{t^n}{ n\,n!}\,, \quad  t \ge 0\,,  \eqno(3)$$
hence the rate of creep is 
$$   
\frac{d \psi^B}{dt}(t) =  
 \frac{\;1-\e^{\ds -  t}}{t} = \sum_{n=0}^{\infty}
   (-1)^{n}  \frac{t^n}{ (n+1)!}\,, \quad t\ge 0 \ \,. \eqno(4)$$
 
  For the \underline{Lomnitz model} \cite{Lomnitz_JG56} we have
  $$ \psi^L(t)= \hbox{log} (1+ t/\tau _0) \,,
  \quad t \ge 0 \,, \quad  \tau _0>0\,,  \eqno(5)$$
where log  denotes the 
{\it natural logarithmic  function}.
Taking again $\tau_0=1$,    
we have the  series representation 
$$ \hbox{log}(1+ t) = 
\sum_{n=1}^{\infty}
   (-1)^{n-1}  \frac{t^n}{ n}\,, \quad  t \ge 0 \,, \eqno(6)$$
which implies a rate of creep
$$   
\frac{d \psi^L}{dt}(t) = \frac{1}{1+t}=  
  \sum_{n=0}^{\infty}
   (-1)^{n}  {t^n}\,,  \quad t\ge 0  \,. \eqno(7)$$
\vspace{0.5cm}   
\begin{figure}
\includegraphics[height=0.90\textwidth,angle=-90]{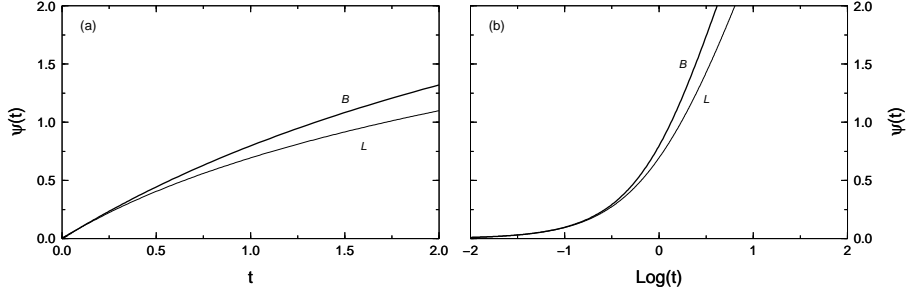}
\vspace{-0.5cm}
\caption{The creep functions for the Becker (B) and Lomnitz (L) rheologies, shown as a function of time.}
\end{figure}

For the Lomnitz law the series representations given by Eqs. (6) and  (7) are convergent only for $0\le t<1$, 
at variance with Eqs. (3) and (4) for the Becker law that are convergent for all $t\ge 0$.
However, all these power series are suitable only for sufficiently small times because their
 {\it numerical} convergence falls down very soon.
 
\vspace{0.5cm}     
 \begin{figure}[h!]
 \includegraphics[height=0.90\textwidth,angle=-90]{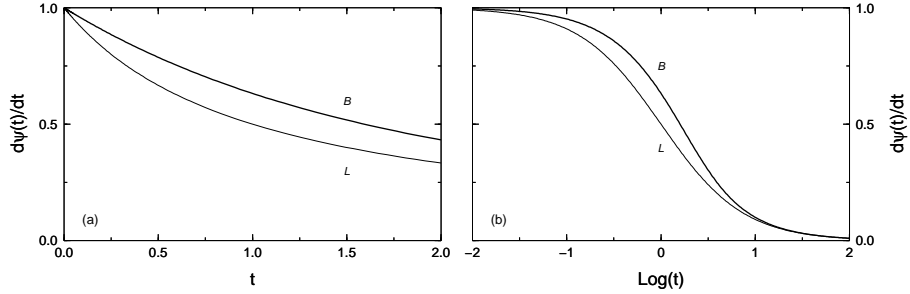}
 \vspace{-0.5cm}
  \caption{Rate of creep for the two models, as a function of time.}
\end{figure}
 In order to compare the creep behavior of the two models,  
 we show $\psi^B(t), \psi^L(t)$ and their time derivatives in Figures 1 and 2, respectively, 
 taking both linear frame (a) and logarithmic (b)  time axes. 
The overall similarity between the two creep functions is apparent in Figure 1,
also showing that the Becker rheology accounts for 
a somewhat larger strain relative to Lomnitz (i.e., $\psi^{B} > \psi^{L}$).
Inspection of the corresponding rates of creep in Figure 2, clearly shows 
that, for finite  values of time,  the 
Becker creep systematically evolves at a larger rate with respect to Lomnitz
(i.e., $d\psi^{B}/dt > d\psi^{L}/dt$). 
For long times, both rates of creep decay to  zero as $1/t$ as we easily note from Eqs. (4) and (7).
      	  
	  \section{The relaxation laws}
The relaxation modulus $G(t)$ for the two rheological models can be derived from the corresponding creep laws
through  the general Volterra integral equation of the second kind \cite{Mainardi_BOOK10}
$$G(t) = \frac{1}{J_U} - \frac{1}{J_U}\, \int_0^t \frac{dJ(t')}{dt'} \,
G(t-t')\, d t'\,.\eqno(8)$$
As a consequence, the dimensionless relaxation function defined by  
$$\phi(t) = J_U \,G(t) \eqno(9)$$
obeys the integral equation
$$\phi(t) = 1 -  q\,\int_0^t\, \frac{d\psi(t')}{dt'}\,
\phi(t-t')\, d t'\,,\eqno(10)$$
where the rate of creep is given by Eqs. (4) and (7) for the Becker and the Lomnitz laws, respectively.
In order to solve numerically Eq. (10), we have used standard numerical methods. 

\vspace{0.5cm}   
  \begin{figure}[h!]
 \includegraphics[height=0.90\textwidth,angle=-90]{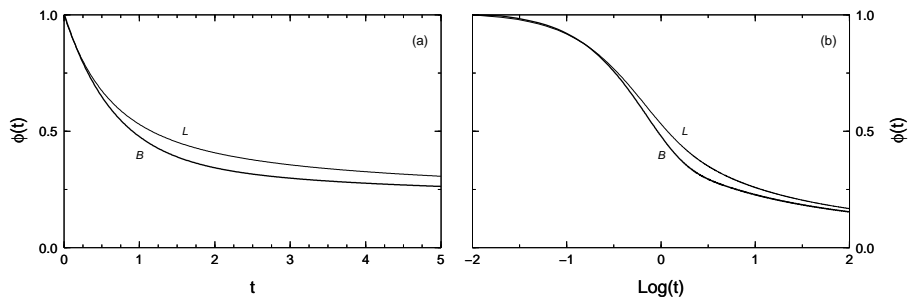}
 \vspace{-0.5cm}
  \caption{Relaxation functions for the two rheological bodies, as a function of time.}
\end{figure}
The results are shown in Figure 3, assuming $q=1$ and adopting again both linear and logarithmic time axes.
As expected according to the similarity of the corresponding creep functions,
 the two relaxation functions show similar features for the two models. 
 However, it is apparent that the Becker model exhibits, at a given time,
  a somewhat larger amount of relaxation relative to Lomnitz
(i.e., $\phi^{B} < \phi^{L}$). 
By visual inspection of the curves it also appears that the Becker rate of relaxation exceeds that
of Lomnitz  model.     
  
	  \section{The retardation spectra}
The determination of the {\it time-spectral functions}  from the knowledge of
the time-dependent material  functions $J(t)$ and $G(t)$ 
is a fundamental problem  from theoretical and experimental view points in polymer science.
It can be formally solved through an analytical  method
outlined by Gross \cite{Gross_BOOK53} based  on the {\it Laplace transform}
of the time derivatives of $J(t)$ and $G(t)$, 
see also \cite{Mainardi_BOOK10}.
For the present models we  use the Gross method  to derive the retardation spectrum 
from the Laplace transform of the rate of creep. 
By definition of the retardation spectrum $R_\epsilon(\tau)$, we have
 $$ J(t) =  J_U +  \displaystyle \int_{0}^{\infty}
  R_\epsilon (\tau)\, \left( 1-\e^{\ds-t/\tau}\right)\, d\tau\,, \eqno(11) $$
where $\tau$ denotes the retardation time.
\newpage
For the two rheological models considered in this note, closed--forms exist for the
retardation spectra. For the Becker model, according to \cite{Gross_BOOK53}, the retardation 
spectrum turns out to be discontinuous, with
     $$ R^B_\epsilon (\tau) = \rec{\tau} H(\tau-1) \,,\eqno(12)$$
	 where $H(t)$ denotes the Heaviside step function, 
while from \cite{Mainardi-Spada_JL12} for the Lomnitz model we have the continuous spectrum
	 $$ R^L_\epsilon (\tau) = \frac{\e^{\ds-1/\tau}}{\tau} \,.\eqno(13)$$   
The two  spectra are compared  in Fig. 4 as function of $\tau$.
 Although these spectra show a dramatic difference character,
 we note that they both show a peak for $\tau=1$ and 
 that they both decay, for $\tau \to \infty$, as $1/\tau$.
  
\vspace{0.5cm}          	  
		  \begin{figure}[h!]
 \includegraphics[height=0.90\textwidth,angle=-90]{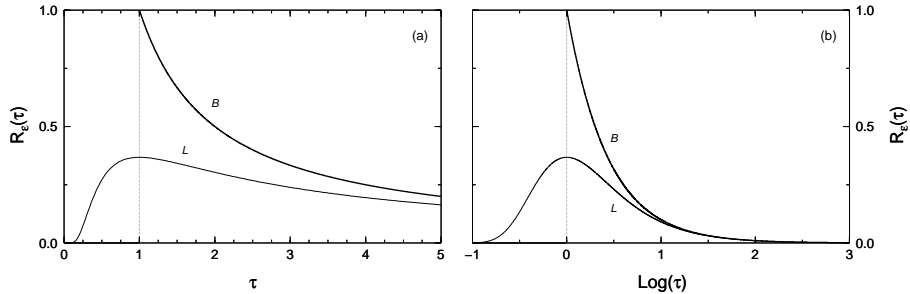}
\vspace{-0.5cm}
  \caption{Retardation spectra for the two rheological models as a function of the retardation time.}
\end{figure}
	  \section {Conclusions}
	 
	 In this paper we have discussed and compared the time-dependent material functions
	 for two viscoelastic models introduced  by Becker and  Lomnitz known
	 in Earth rheology. These functions for the two models show a broadly similar
	 behavior, so that they could hardly be discriminated from experimental point of view.
	     Despite this similarity, the corresponding retardation spectra show a dramatic difference.
		 While the Lomnitz spectrum varies smoothly on the whole range of retardation times,
		 the Becker one displays a cut off for short time even allowing a similar decay at large times.
		The examples discussed in this paper clearly show that in order to discriminate 
		between two rheological models exhibiting very similar creep and relaxation behaviors
		the evaluation of the spectra is required.

\end{document}